\documentclass[12pt]{JHEP} 
\usepackage{graphicx}

\newcommand\req[1]{Eq.~(\ref{#1})}
\newcommand\rab[1]{Table~\ref{#1}}

\newcommand\im{\mathrm{Im}}
\newcommand\gev{\mathrm{GeV}}
\newcommand\si{S_{\rm I}}
\newcommand\ti{T_{\rm I}}
\newcommand\sr{S_{\rm R}}
\newcommand\tr{T_{\rm R}}
\newcommand\fig[1]{Fig.~\ref{#1}}
\newcommand\dpar[2]{\frac{\partial #1}{\partial #2}}

\def\be{\begin{equation}}
\def\ee{\end{equation}}
\def\br{\begin{eqnarray}}
\def\er{\end{eqnarray}}

\title{Moduli as Inflatons in Heterotic M-theory}
\author{T. Barreiro\thanks{Work supported by PPARC.} \\
Centre for Theoretical Physics, University of Sussex  \\
Falmer, Brighton BN1 9QJ, UK \\ \email{mppg6@pcss.maps.susx.ac.uk}} 

\author{B. de Carlos \\
Theory Division, CERN, CH-1211 Geneva 23, Switzerland \\ 
\email{Beatriz.de.Carlos@cern.ch}}

\abstract{
We consider different cosmological aspects of heterotic M-theory. In 
particular we look at the dynamical behaviour of the two relevant 
moduli in the theory, namely the length of the eleventh segment 
($\pi \rho$) and the volume of the internal six manifold ($V$) in
models where supersymmetry is broken by multiple gaugino condensation.
We look at different ways to stabilize these moduli, namely racetrack 
scenarios with and without non-perturbative corrections to the K\"ahler 
potential. The existence of different flat directions in the scalar 
potential, and the way in which they can be partially lifted, is discussed 
as well as their possible role in constructing a viable model of inflation. 
Some other implications such as the status of the moduli problem within 
these models are also studied.}

\preprint{CERN-TH/99-401\\
          SUSX-TH-99-019}

\keywords{M-Theory, Supersymmetry Breaking, Cosmology of Theories beyond 
the SM}

\begin{document}
\section{Introduction}

The formulation by Ho\v{r}ava and Witten \cite{Horav96} of the field 
theoretical limit of strongly coupled string theory (or M-theory) as 
an eleven-dimensional ($d=11$) supergravity (SUGRA) compactified on a 
manifold with boundaries (explicitly studied in the case of a Calabi--Yau 
(CY) manifold times the eleventh segment \cite{Witte96}), coupled to $d=10$ 
supersymmetric (SUSY) Yang--Mills theories, opened up new ways of trying 
to understand the different problems that we have come across in our 
attempt to use string theories to obtain a complete picture of 
the elementary particle world. 

In particular we have in this theory two components of the boundary
which contain one $E_8$ super-Yang--Mills sector each. They communicate
through gravitational interactions and one of them is more strongly 
coupled than the other. Altogether the picture we get is that of two 
walls (so-called hidden and observable) that interact through gravity, 
the more strongly coupled being a straightforward candidate to give rise 
to the condensation of gauginos \cite{Horav96p}, the fermionic partners of 
gauge bosons.

Among the different phenomenological achievements of these constructions 
we find that the experimentally determined unification scale of $10^{16}$ 
GeV can be reconciled with the effective $d=4$ Planck scale of $10^{18}$ GeV 
\cite{Witte96,Banks96}, the pattern of soft breaking terms can give gaugino 
masses of the order of scalar masses \cite{Nilles97}, and there are 
candidates for the QCD axion \cite{Choi97} and quintessence \cite{Choi99}.

However, and despite all these successes, the two main issues concerning 
string theories in general remain unsolved, namely those of moduli 
stabilization and supersymmetry breaking. Whether these two are related
or not is still an open question (see, for example, the recent discussion
of \cite{Dine99}) but, for the remainder of this paper, we shall assume 
that they are. More precisely, we shall study the stabilization of the 
moduli (i.e. the dilaton $S$ and modulus $T$) through gaugino 
condensation in the hidden wall. These moduli are the superfields whose 
vacuum expectation values (VEVs) are directly related to two observable 
quantities, the volume of the CY manifold and the length of the eleventh 
segment. In order to reproduce the phenomenologically preferred values 
for the unification scale and coupling constant, those two quantities must 
be of order $(3 \times 10^{16} \; {\rm GeV})^{-6}$ and 
$(4 \times 10^{15} \; {\rm GeV})^{-1}$ respectively.
  
Therefore, given the scales at which gauginos condense ($\sim 10^{11}$ 
GeV), we will work directly with the $d=4$ effective theory coming from 
$d=11$ SUGRA, disregarding an intermediate $d=5$ effective theory in between
the  Calabi--Yau compactification scale, and that of the eleventh segment 
\cite{Banks96,Ellis99}. Altogether, we follow the approach of Choi and
collaborators \cite{Choi98} by also considering the effect of membrane 
instantons that result in a non-perturbative contribution to the K\"ahler 
potential. 

Once we have defined a model within which we achieve SUSY breaking at
the right scale, we can proceed to study the cosmological behaviour of 
these moduli fields. Inflation due to F-terms in the scalar potential has 
always been considered a generic problem of SUGRA theories essentially due
to the absence of flat-enough directions. However several attempts have
been made to build string-inspired inflationary potentials, having 
either the moduli \cite{Binet86,Copel94} or matter fields \cite{Casas97} 
as inflatons. 

In our case we start our analysis of the cosmological viability of
our model by noting very general features of the scalar potential defined 
by multiple gaugino condensation (or racetrack) models in the context of 
M-theory, namely the existence of flat and almost flat directions
for two and more condensates, respectively. As mentioned before, 
flat directions are always of interest in cosmology, and indeed we study 
the capability of these models to drive inflation and, in particular, of 
the moduli fields to become suitable inflatons. 

Together with inflation, the other main cosmological issue about 
string-derived models is the so-called moduli problem \cite{Decar93p}.
It turns out that, in general, moduli are very weakly interacting particles
with VEVs of the order of the Planck scale and masses of the order of
the electroweak scale, which makes them very dangerous relics from the
point of view of the evolution of the Universe. In the final part of this
paper we consider this problem, calculating their masses and pointing out 
which of them would be in a cosmologically dangerous range.

Therefore the plan of this paper is as follows. In section~2 we study the 
stabilization of the moduli through gaugino condensation in the hidden
wall. First of all we point out a few general features of scalar potentials
due to multiple gaugino condensation, which are independent of the
form of the K\"ahler potential used, namely the presence of flat and
almost flat directions. We shall also see that, in order to achieve the 
desired VEVs for the moduli, with SUSY broken at the right scale (given by 
a gravitino mass, $m_{3/2}$, of about $1$ TeV) we can either invoke pure
multiple gaugino condensation (also denoted as the racetrack mechanism) or
combine it with some non-perturbative corrections to the K\"ahler potential, 
which will ensure the cancellation of the cosmological constant at the 
minimum (the so-called K\"ahler stabilization). In order to illustrate 
these results we present a detailed calculation with a particular ansatz 
for $K_{\rm np}$. 

In section~3 we consider the cosmological evolution of these fields. 
The presence of the flat directions demonstrated in section~2 opens up the 
possibility of achieving inflation with some combination of the moduli
fields as the inflaton. We study the behaviour of the different candidates 
for inflatons, and we select the ones that give an acceptable pattern of
inflation. Continuing with cosmological issues, in section~4 we study the 
moduli problem of these models. We calculate the moduli masses, and point 
out possible problems of those associated with the flat directions, as well 
as possible solutions. We conclude in section~5.

\section{Stabilization of the moduli}

As mentioned in the introduction, we want to study various phenomenological
features of compactified heterotic M-theory constructions where SUSY is 
broken by gaugino condensation. The two dynamical quantities we need to 
stabilize are $\rho$, the length of the eleventh segment, and $V$, the
volume of the internal manifold. Those are related to the chiral
superfields $S$, $T$ through
\br
{\rm Re} \; S & = & (4 \pi)^{-2/3} \kappa^{-4/3} V \;\; , \nonumber \\
&  & \label{fields} \\ 
{\rm Re} \; T & = &  \left( 4 \pi \sum_{IJK} C_{IJK}/6 \right)^{-1/3} 
\kappa^{-2/3} \pi \rho V^{1/3}  \;\; , \nonumber
\er
where $\kappa^2$ is the $d=11$ gravitational coupling, and $C_{IJK}$
are the CY intersection numbers.
 
From now on we shall follow the notation of Ref.~\cite{Choi98} and define
the phenomenologically preferred values:
\br
\langle {\rm Re} \; S \rangle & = & \langle S_{\rm R} \rangle = 
O(\alpha_{\rm GUT}^{-1}) \;\; , \nonumber \\
&  & \label{app} \\
\langle {\rm Re} \; T \rangle & = &\langle T_{\rm R} \rangle = 
O(\alpha_{\rm GUT}^{-1}) \nonumber \;\; ,
\er
in units of the $d=4$ Planck mass, $M_{\rm P}$ ($=\kappa^{-1} 
\sqrt{\pi \rho V}$). In order to study the stabilization of these fields 
we need to calculate the scalar potential, which is made out of two 
functions, the superpotential $W$ and the K\"ahler potential $K$. The latter 
can be divided into two pieces, a perturbative one ($K_{\rm p}$), which 
admits an expansion in powers of $1/S_{\rm R}$, $1/T_{\rm R}$ (and can be 
determined in different M-theory limits \cite{Li97,Lukas98}), and 
$K_{\rm np}$, which contains all non-perturbative effects. Therefore we have
\be
K = K_{\rm p} + K_{\rm np} = K_0 + \delta K_{\rm p} + K_{\rm np} \;\;,
\label{kahler}
\ee
where
\be
K_0 = -\log(S+\bar{S}) - 3 \log(T+\bar{T})
\label{treekahler}
\ee
and $\delta K_{\rm p}$ stands for higher-order corrections to the 
perturbative expansion. In this context, $K_{\rm np}$ stands for the 
M-theory version of stringy non-perturbative effects \cite{Shenk90}, 
which are given by different types of instantons wrapping several cycles 
of the CY. These were computed first in \cite{Becke95} for the case of 
type IIA M-theory and the discussion was extended in \cite{Choi98} to 
heterotic M-theory. Out of the three different types of instantons we 
can have, membrane instantons wrapping the CY 3-cycle ($I_1 \sim b 
\sqrt{S_{\rm R}}$), membrane instantons wrapping the CY 2-cycle ($I_2 \sim 2 
\pi T$), and five-brane instantons wrapping the entire CY volume ($I_3 
\sim 2 \pi S$), only the first ones are going to give a significant 
contribution to $K_{\rm np}$, which goes as $e^{-I_1}$.

As for the superpotential, we shall consider multiple gaugino condensation
\be
W = \sum_a C_a e^{-\alpha_a f_a} + O(e^{-2 \pi S}, e^{-2 \pi T}) \;\; ,
\label{sup}
\ee
where $a$ labels the corresponding hidden gauge group $G_a$, $\alpha_a$ 
are the corresponding one-loop beta function coefficients ($\alpha_a = 24 
\pi^2/(3 N_a-M_a)$ for SU($N_a$) with $M_a$ pairs of matter fields 
transforming as $N_a$, $\bar{N}_a$), the $C_a$ coefficients
depend on $N_a$ and $M_a$,
and $f_a$ are the gauge kinetic functions, defined as
\be
f_a = \frac{1}{4 \pi} \left( S - \frac{n_a}{2} T \right) + O(e^{-2 \pi S}, 
e^{-2 \pi T}) \;\;.
\label{fkin}
\ee
Their exact form is currently known for smooth CY compactifications
\cite{Choi97,Nille97,Lukas98}, even in the presence of five-branes 
\cite{Benak99}, and they have been studied for orbifold compactifications 
in \cite{Stieb99}\footnote{For most of the examples studied so far the 
threshold corrections to the gauge kinetic functions are known to be 
universal, i.e. $n_a$ are the same for any group in the hidden 
wall. However there exist particular examples, such as enhanced gauge
group thresholds, which may yield different $n_a$ for the different 
condensing groups.}.

Finally the scalar potential is given by
\be
V = e^K \left\{ \left| W_S + K_S W \right|^2 (K_{S \bar{S}})^{-1}
    + \left| W_T + K_T W \right|^2 (K_{T \bar{T}})^{-1}
    - 3 \left| W \right|^2 \right\} \;\; ,
\label{pot}
\ee
where the subindices indicate derivatives of $W$ and $K$ with respect to the
different fields.

The next step will be to minimize this potential, and study the possibility
of finding minima with zero cosmological constant, i.e.\ we want the 
conditions $\partial V/\partial S = \partial V/\partial T =0$ (together 
with $V|_{\rm min} = 0$ whenever it is possible) to be fulfilled for values 
of $S_{\rm R}$, $T_{\rm R}$ given by Eq.~(\ref{app}). Before getting into 
the numerical details of our calculation, let us note a very distinctive 
feature of these kinds of potentials, namely the possibility of having flat 
directions along the imaginary components of $S$ and $T$. This fact was 
already noted in \cite{Choi98} for the case of two condensates, and is 
generalized here, independently of the form of $K$. It only relies on the 
form of the gauge kinetic functions $f_a$.

\subsection{Imaginary directions}\label{imdir}

We will start our analysis by looking for minima in the imaginary 
directions of the moduli fields, $\si = \im \; S$ and $\ti = \im \; T$. 
In the case of one single condensate, where the superpotential is 
$W = C e^{-\alpha (S - n T/2)}$, the scalar potential of \req{pot} is totally 
independent of the phase of the superpotential. Therefore both $\si$ and 
$\ti$ will be flat directions of the scalar potential.

Let us now consider the case of a two-condensate racetrack model, where the
superpotential of \req{sup} is given by
\be
W = W_1 + W_2 \;\; ,
\label{sup2c}
\ee
with $W_1 = C_1 e^{-\alpha_1 (S - n_1 T / 2)}$ and
$W_2 = C_2 e^{-\alpha_2 (S - n_2 T/2)}$. Intuitively, one expects that one 
of the phases can still be factored out from the scalar potential, as in 
the one-condensate case, leaving a flat direction in the potential. We can 
easily check this result by calculating the derivatives of the scalar 
potential with respect to the imaginary fields. From \req{pot}, we obtain 
the derivative with respect to $\si$:
\be
\frac{\partial V}{\partial \si} = 
-2 \ \im \left( {\frac{\partial V}{\partial S}} \right) = 
- \frac{\alpha_2 - \alpha_1}{2 \pi}\; e^{K} A \ \im (\bar{W_1} W_2) \;\;,
\label{dvdsi}
\ee
where 
\be
A = 3 - \frac{(K_{\bar{S}} - \alpha_1/4\pi) (K_S - \alpha_2/4\pi)}
{K_{S \bar{S}}} - \frac{(K_{\bar{T}} + n_1 \alpha_1/8\pi) (K_T + n_2 
\alpha_2/8\pi)}{K_{T \bar{T}}} 
\label{Acst}
\ee
is a quantity independent of $\si$ and $\ti$. Similarly, the derivative with 
respect to $\ti$ is
\be
\frac{\partial V}{\partial \ti} = 
-2 \ \im \left( {\frac{\partial V}{\partial T}} \right) = 
 \frac{n_2 \alpha_2 - n_1 \alpha_1}{4 \pi} e^{K} A \ \im (\bar{W_1} W_2) \;\;,
\label{dvdti}
\ee
with $A$ still given by \req{Acst}. A sufficient condition for a stationary 
point is then
\be
\im ( \bar{W_1} W_2 ) = 0 \;\;,
\label{necmin}
\ee
which  yields
\be
\Phi^-_{\rm i} \equiv \si - \frac{n_1 \alpha_1 - n_2 \alpha_2}{2(\alpha_1 - 
\alpha_2)} \ti = \frac{4 \pi^2 k }{\alpha_1 - \alpha_2} \;\;,
\label{mincond}
\ee
where we have defined a new field $\Phi^-_{\rm i}$ and $k$ is an integer. 
It is easy to check that $k$ even corresponds to a maximum and $k$ odd to a 
minimum. More importantly, this relation does not fix both $\si$ and $\ti$, 
a good indication that there is a flat direction in the potential. Indeed, 
the direction orthogonal to $\Phi^-_{\rm i}$, 
\be
\Phi^+_{\rm i} \equiv \frac{n_1 \alpha_1 - n_2 \alpha_2}{2(\alpha_1 - 
\alpha_2)} \si + \ti \;\; ,
\label{imflat}
\ee
is a flat direction of the potential, for which $\dpar{V}{\Phi^+_{\rm i}} = 
0$ for all values of the fields.

So what happens if we have more than two condensates? Naively we would
expect that, even though one of the phases can be factored out from 
the scalar potential as before, the remaining ones should be able to 
stabilize both directions in the imaginary fields. The superpotential is 
given by the general formula of \req{sup}, and the imaginary derivatives 
will be
\be
\dpar{V}{\si} = -\frac{e^K}{2 \pi} \sum_{i<j} \left[ A_{ij} (\alpha_j -
\alpha_i) \;   \im( \bar{W_i} W_j ) \right] \;,
\label{ndvdsi}
\ee
and 
\be
\dpar{V}{\ti} = \frac{e^K}{4 \pi} \sum_{i<j} \left[ A_{ij} (n_j \alpha_j - n_i
\alpha_i)  \;  \im( \bar{W_i} W_j ) \right] \;\;,
\label{ndvdti}
\ee
where
\be
A_{ij} = 3 - \frac{(K_S - \alpha_i/4 \pi)(K_{\bar{S}} - \alpha_j/4 \pi)}
{K_{S \bar{S}}}
 - \frac{(K_T + n_i \alpha_i/8 \pi)(K_{\bar T} + n_j \alpha_j/8 \pi)}
{K_{T \bar{T}}}
\label{An}
\ee
are again independent of $\si$ and $\ti$.

The condition for the existence of a flat direction, 
$a \dpar{V}{\si} + b \dpar{V}{\ti} = 0$ for some $a$ and $b$ independent 
of $\si$ and $\ti$, becomes
\be
0 = \sum_{i<j} \left[ A_{ij} \left( a (\alpha_j - \alpha_i) 
+ \frac{b}{2} (n_j \alpha_j - n_i \alpha_i) \right) \; \im( \bar{W_i} W_j ) 
\right] \;.
\label{condn}
\ee
In the case of a single gauge kinetic function parameter, $n = n_i = n_j$ 
for all $(i,j)$, and $ a = -n b/2$ is then an obvious solution.  In other 
words, if all the $n_i$ are the same, there is always a flat direction in 
the imaginary fields, {\em for any number of condensates in the hidden 
sector}. On the other hand, if one of the $n_i$ differs from the others, it 
is impossible to find a solution to \req{condn} holding for all $\si$ and 
$\ti$, so all the flat directions in the imaginary fields become lifted.

\FIGURE{
\parbox{15cm}{
\includegraphics[width=7.5cm]{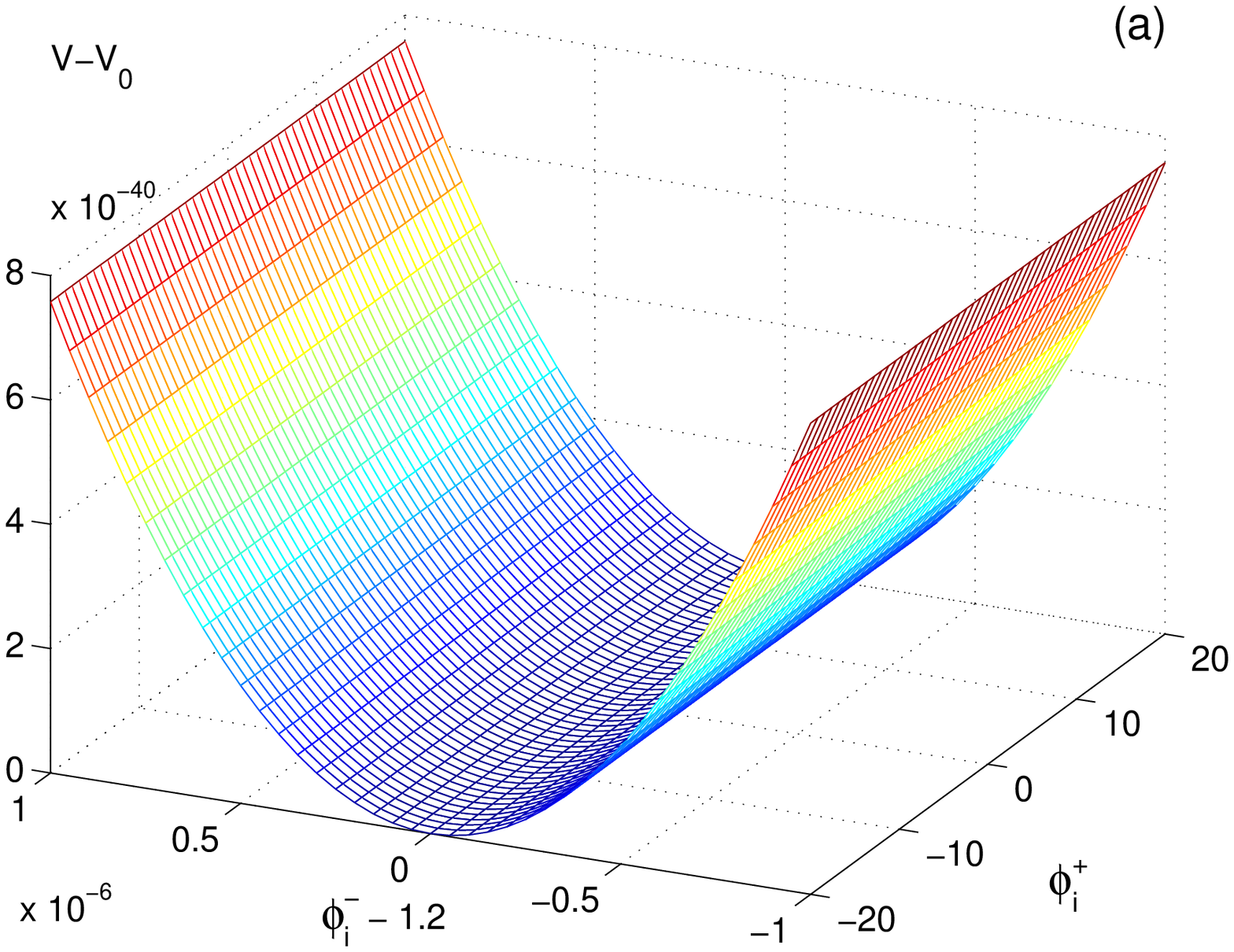}
\includegraphics[width=7.5cm]{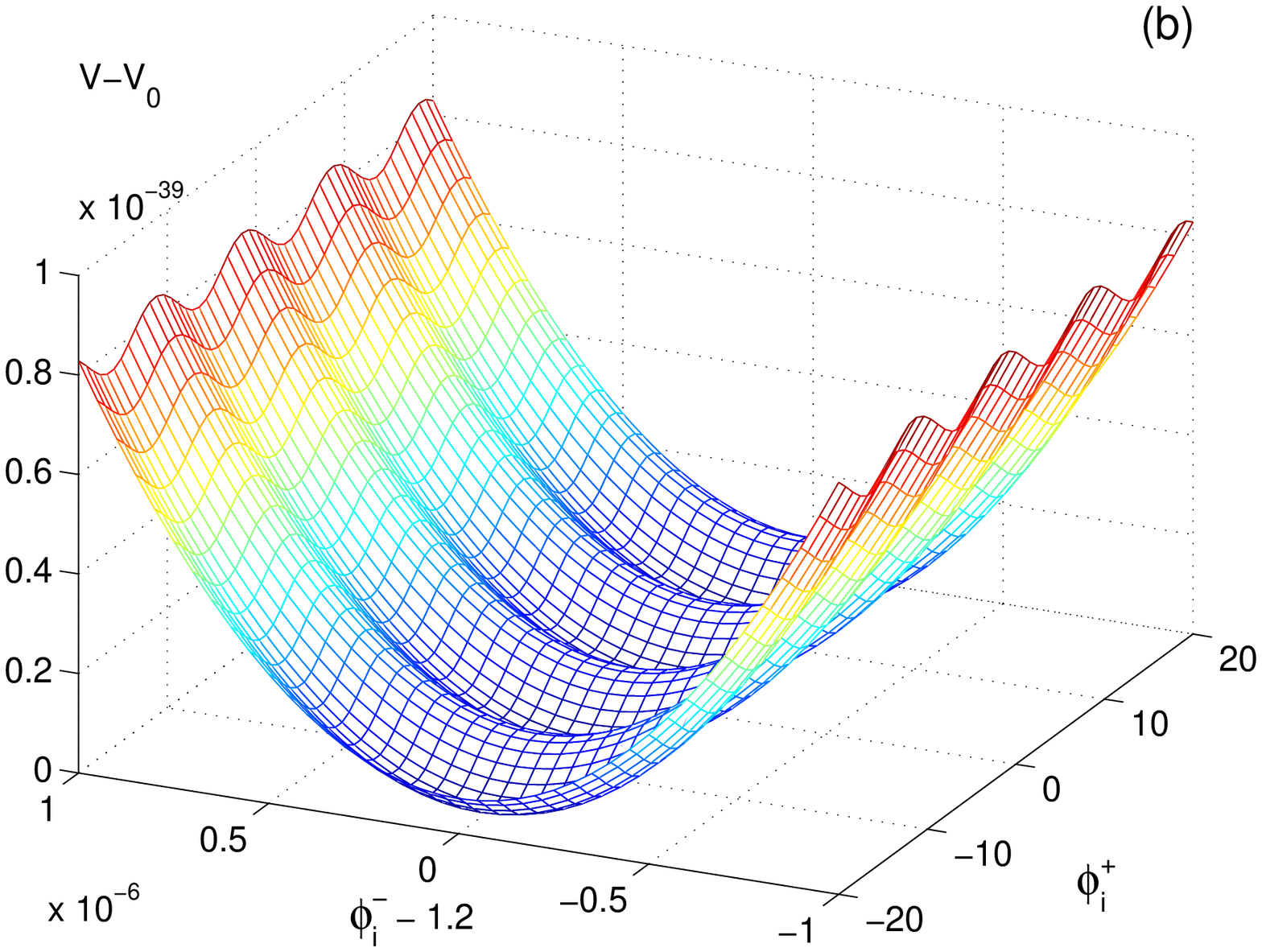}
\caption{3-d plot of the scalar potential $V$ as a 
function of $\Phi^-_{\rm i}=\si-\ti/2$ and $\Phi^+_{\rm i}=\si/2+\ti$ 
(in $M_{\rm P}$ units) for the case of \textbf{\sffamily (a)}
two condensates with 
$N_1=3$, $M_1=0$, $N_2=4$, $M_2=8$ and gauge kinetic function parameters 
$n_1 = n_2 = 1$ and \textbf{\sffamily (b)} three condensates, consisting of the previous 
two plus a third one given by $N_3=2$, $M_3=0$ and $n_3=0.5$. We can see 
that the flat direction of {\sffamily (a)} is lifted in {\sffamily (b)}
by the presence 
of the third condensate. In both examples we have set the $K_{\rm np}$ 
parameters to $D = 0.13$ and $B = 120$, and fixed the real values of the 
fields to their minima $\sr \sim 19$ and $\tr \sim 15.9$ (see section 
\ref{realdir}). $V_0$ is the value of the potential at the minimum.
}
\label{fig1}
}
}

As an illustration of this vacuum structure for these models, we have 
plotted in \fig{fig1} the scalar potential versus the orthogonal 
combinations of the imaginary fields, for an example of two and 
three condensates. We can clearly see in the second picture how the flat 
direction is lifted by the presence of the third condensate {\em with 
a different gauge kinetic function parameter}. We will explore the 
cosmological interest of these features in section~\ref{cosmevol}.

\subsection{Real directions}\label{realdir}
     
We will look now at the minimization of the potential, \req{pot}, along the 
real components, $\sr$ and $\tr$. In particular we are going to analyse 
two different situations: K\"ahler stabilization and pure racetrack models. 

Let us start with the pure racetrack case, that is, a K\"ahler potential 
given only by its perturbative piece, $K = K_0$ in \req{kahler}. The real 
directions are in a way similar to the imaginary directions, in the sense 
that the interplay between two condensates, through the superpotential
contributions to the scalar potential of \req{pot}, will be able to fix
the combination $\Phi_{\rm r}^{-} = \sr - 
\frac{n_1 \alpha_1 - n_2 \alpha_2}{2 (\alpha_1 - \alpha_2)} \tr$, whereas 
the orthogonal combination $\Phi^+_{\rm r} = 
\frac{n_1 \alpha_1 - n_2 \alpha_2}{2 (\alpha_1 - \alpha_2)} \sr + \tr$ will
factor out from the superpotential contributions. However, contrary to the 
imaginary case, this direction will not be flat because of extra terms 
arising from the K\"ahler potential contributions to \req{pot}. We have 
checked that these K\"ahler terms will always yield a runaway potential, 
provided all the gauge kinetic parameters $n_i$ are the same. This is true 
for any number of condensates in the hidden sector. So if we want to find 
a minimum for the real fields, we need to introduce different $n_i$ or
non-perturbative corrections to the K\"ahler potential.

To illustrate the first case, we have found a few examples where the real 
directions are indeed fixed with only two condensates having different 
gauge kinetic function parameters. These are presented in \rab{exampin}.
The results are similar to the usual ones obtained in the weakly coupled 
case \cite{Decar93}, in particular the potential energy at the minimum 
turns out to be always negative.
\TABLE{
\begin{tabular}{|ccc|ccc|cc|c|}
\hline
$N_1$ & $M_1$ & $n_1$ & $N_2$ & $M_2$ & $n_2$ & $\sr$ & $\tr$ & 
$m_{3/2} (\gev)$ \\ \hline
3     &     8 &  1.00 &   2   &   0   &  1.06 & 17.6  & 20.7  &  1090      \\
3     &     7 &  1.40 &   2   &   0   &  1.45 & 22.0  & 20.3  &   120      \\
 5    &     7 &  1.00 &   4   &   0   &  1.04 & 26.9  & 23.7  &    74      \\
\hline
\end{tabular}
\caption{Some examples of minima in a pure racetrack scenario.}\label{exampin}
}

We now turn to the second possibility, including non-perturbative
corrections to the K\"ahler potential. In order to do that we choose a 
particular ansatz for $K_{\rm np}$ \cite{Barre98}, which gives a zero 
cosmological constant at the minimum of the potential,
\be
K_{\rm np} = \frac{D}{B\sqrt{\sr}} 
\log\left( 1+e^{-B (\sqrt{\sr}-\sqrt{S_0})} \right) \;\;,
\label{ansatz}
\ee
where we fix $S_0=19$, and the different vacua, for a given $W$, are 
functions of pairs of values ($D$, $B$).

Apart from having a different ansatz for $K_{\rm np}$, the rest of the vacuum
structure is totally analogous to that found in \cite{Choi98}. The case of
one condensate does not result in any minima with the right order of
magnitude for $\sr$, $\tr$ (see Eq.~(\ref{app})), whereas two or more 
condensates yield the desired results, with the $K_{\rm np}$ correction
stabilizing the previously runaway direction $\Phi_{\rm r}^+$. To obtain
reasonable values for $\sr$ and $\tr$, we require the presence of hidden 
matter fields, transforming under the hidden gauge groups, in a way similar
to what happened in weakly coupled heterotic string theories \cite{Decar93}.

We will now consider the cosmological possibilities of both
these types of solutions in turn.

\section{Cosmological evolution}\label{cosmevol}

In order to study the cosmological evolution of the moduli fields we
have to solve their equations of motion in an expanding Universe. These
are given by
\br
K_{S \bar{S}} ( \ddot{S} + 3 H \dot{S} ) + K_{S S \bar{S}} \dot{S}^2 + 
\frac{\partial{V}}{\partial{\bar{S}}} & = & 0 \nonumber \\ 
& & \label{evol} \\
K_{T \bar{T}} ( \ddot{T} + 3 H \dot{T} ) + K_{T T \bar{T}} \dot{T}^2 + 
\frac{\partial{V}}{\partial{\bar{T}}} & = & 0 \;\;, \nonumber
\er
with $H$, the Hubble constant, given by
\be
H^2 = \frac{1}{3} K_{S \bar{S}} \dot{S} \dot{\bar{S}} + \frac{1}{3} 
K_{T \bar{T}} \dot{T} \dot{\bar{T}} + \frac{V}{3} \;.
\label{H}
\ee
As we can see, the presence of non-canonical kinetic terms for $S$ and $T$
already imposes a few modifications in the above system, with respect
to the usual canonically normalized fields. We will
consider in turn the cases of two and three condensates.

\subsubsection*{Two condensates}
                       
Summarizing what was found in the previous section, we have four
possible directions of the scalar potential, which are functions of
the two complex moduli fields $S$ and $T$. Those corresponding to the 
combinations $\Phi^-_{\rm r,i}$ are, as we said before, fixed by the 
interplay between the two condensates,  much as was done in 
the weakly coupled heterotic string \cite{Decar93}. Therefore, along these 
two directions, the behaviour from the cosmological point of view is very 
similar to the one in the stringy case, which was studied in 
Ref.~\cite{Brust93}, only that now the gauge kinetic functions are given by 
Eq.~(\ref{fkin}) rather than just $4 \pi f_a=k_a S$ ($k_a$ is the Kac--Moody 
level of the corresponding gauge group). The conclusion of that study was 
that, along the $S$ direction (which is $\Phi^-$ now) the potential is too 
steep to the left of the minimum to guarantee that the field will settle 
down in the latter instead of rolling past. Moreover its kinetic energy is 
too big to drive inflation during its evolution.

We face here exactly the same situation and therefore conclude that,
along the above-mentioned directions, no new features arise. However it
is worth mentioning that, in the presence of a dominating background,
this combination of fields could reach a scaling regime just like
$S$ in the heterotic string case \cite{Barre98p}. The effects 
of such background in these M-theory models will be studied elsewhere
\cite{Barre99}.

\FIGURE{
\parbox{15cm}{
\includegraphics[width=7.5cm]{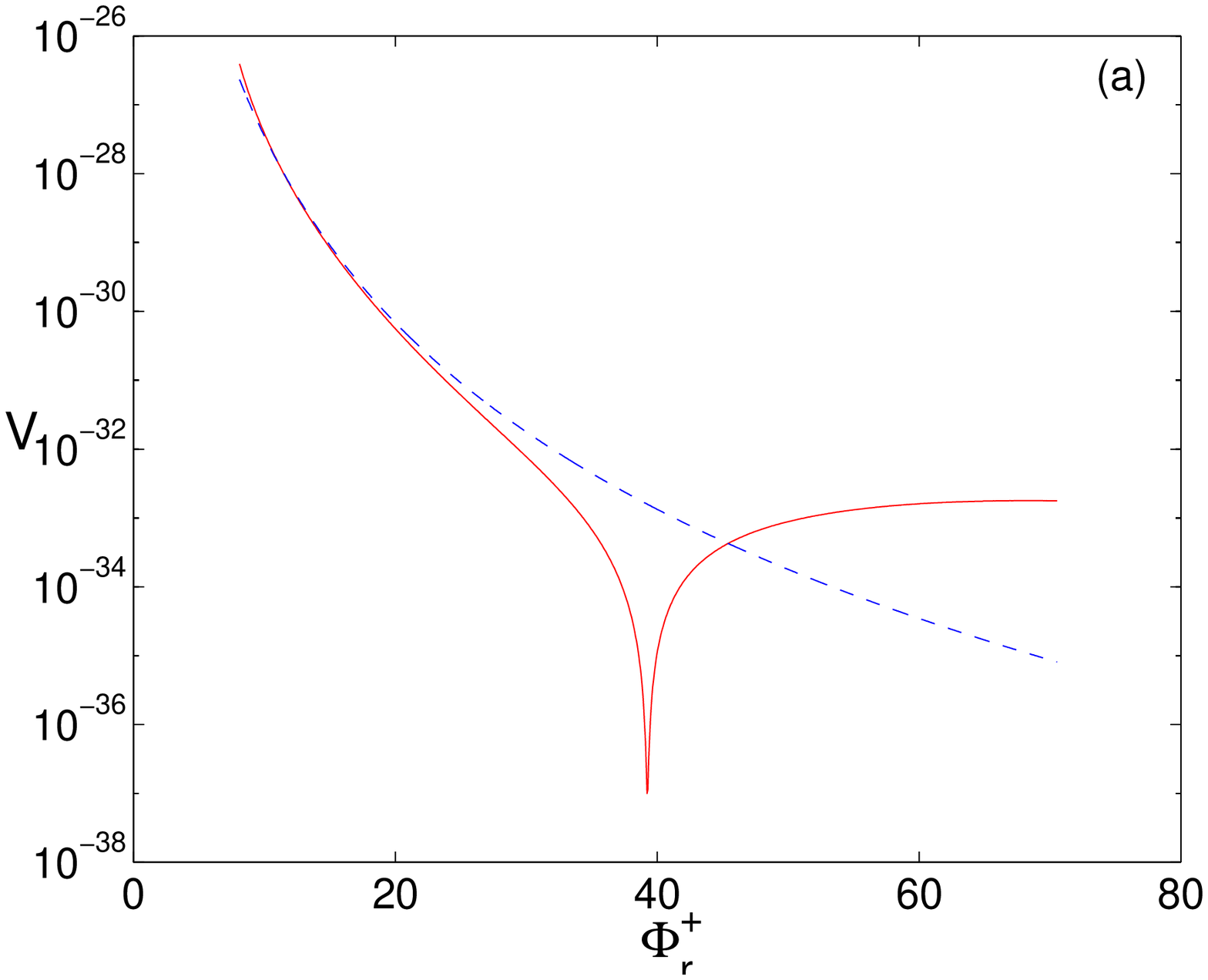}
\includegraphics[width=7.5cm]{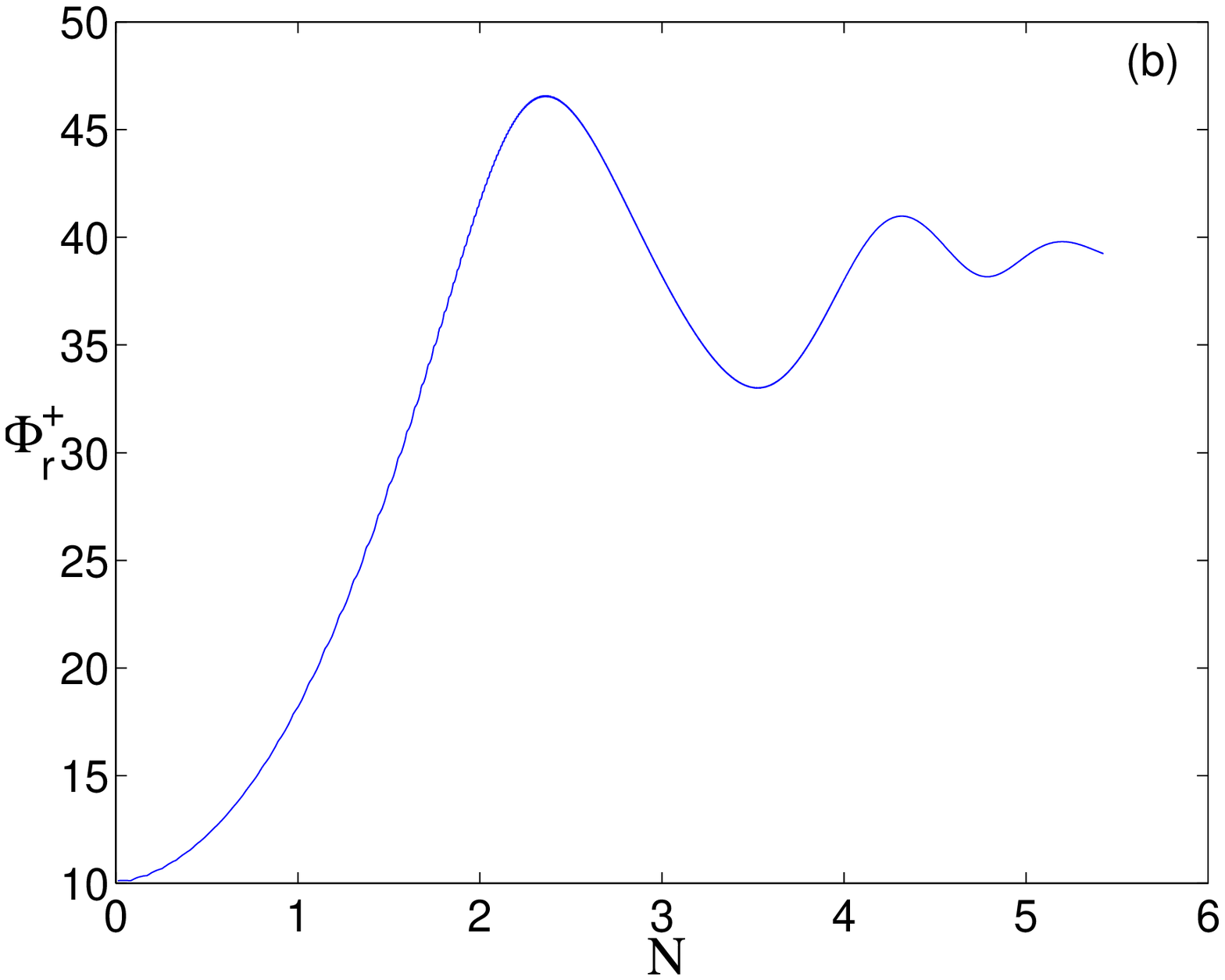}
\caption{ \textbf{\sffamily (a)} Solid: $V$ as a function of 
$\Phi_{\rm r}^+$ (in $M_{\rm P}$ units) along the valley defined by 
$\Phi^-_{\rm r}=11.07$ for the same example as in Fig.~1a. Dashed: a 
simple approximation to $V$ given by $Q/(\Phi^+_{\rm r})^9$;
\textbf{\sffamily (b)} 
evolution of $\Phi^+_{\rm r}$ as a function of $N$ (the number of e-folds) 
along the same valley as before, but with $K_{np}$ parameters $B = 3$ and
$D = 13.9924$.}
\label{fig2}
}
}

Let us now turn to the analysis of the other two remaining directions, 
namely $\Phi^+_{\rm r,i}$. As mentioned before, $\Phi^+_{\rm i}$ is a 
totally flat direction in the case of two condensates, and therefore no 
evolution is expected. At this stage we could consider the effect of 
including the higher-order corrections indicated in Eqs.~(\ref{sup}) and 
(\ref{fkin}), given that they would introduce terms depending 
on $\si$ and $\ti$. However, when we took these terms into account, we 
found that the numerical effect is minute, and that no significant 
differences with respect to the totally flat case can be seen. Nevertheless 
these corrections can give rise to very interesting potentials concerning 
the axion mechanism to solve the strong CP problem \cite{Choi97} or 
quintessence \cite{Choi99}. Neither is it worth considering higher order 
corrections to the K\"ahler potential, such as those parametrized by 
$\delta K_{\rm p}$ in Eq.~(\ref{kahler}), given that this function 
carries no dependence on the imaginary parts of either $S$ or $T$. 

We are then left with the evolution along $\Phi^+_{\rm r}$. This is, {\em 
a priori}, the most promising direction along which to study the evolution 
of the fields, given that, as mentioned in section~\ref{realdir}, this 
field only gets contributions to its scalar potential from the K\"ahler 
potential. That is, instead of having an exponential potential like 
$\Phi_{\rm r}^-$, it has a power-law type one. This is true for both the 
pure racetrack models, where $K_{\rm np} = 0$, and the K\"ahler 
stabilization models. An example of the evolution of the $\Phi^+_{\rm r}$ 
for the latter models is shown in Fig.~2 for the same two condensing 
groups as in Fig.~1a, but with $K_{\rm np}$ defined by $B=3$, $D \sim 14$, 
this time. In this example the direction shown is the orthogonal to that 
defined by $\Phi^-_{\rm r} \sim 11.07$, and the imaginary fields were fixed 
to their minimum values. In Fig.~2a we have plotted the scalar potential 
as a function of $\Phi_{\rm r}^+$, as well as a quite accurate fit to its 
slope given by $Q/(\Phi_{\rm r}^+)^9$, with $Q \sim 3.5 \times 10^{-19}$.
Concerning the cosmological behaviour, we have evolved both 
$\Phi_{\rm r}^+$ and $\Phi^-_{\rm r}$ simultaneously, in order to 
take into account the small oscillations that the latter has around its 
central value, defining a `valley' in the potential as both fields evolve 
towards the minimum. As we see in Fig.~2b, a maximum of $\sim 5$ e-folds of 
evolution can be obtained before the field reaches its minimum (defined 
by $\sr \sim 24.5$, $\tr \sim 27$, i.e. $\Phi_{\rm r}^+=39.25$) and 
oscillates around it. Even though this is not a very promising candidate 
for an inflaton, it is interesting to note the fact that the field does 
stay at its minimum for a big fraction of initial conditions within this 
valley. 

In fact, these kinds of inflationary models (defined by inverse power-law 
potentials) are denoted in the literature as `intermediate inflation' 
\cite{Musli90}, and the reason why, in this case, the inflationary period 
does not last for long enough is that neither $S$ nor $T$ have canonical 
kinetic terms. With non-minimal kinetic terms the evolution is considerably 
faster than that of a typical intermediate inflation model. Note that this 
is also true for the pure racetrack models.

\subsubsection*{Three condensates}

\FIGURE{
\parbox{15cm}{
\includegraphics[width=7.5cm]{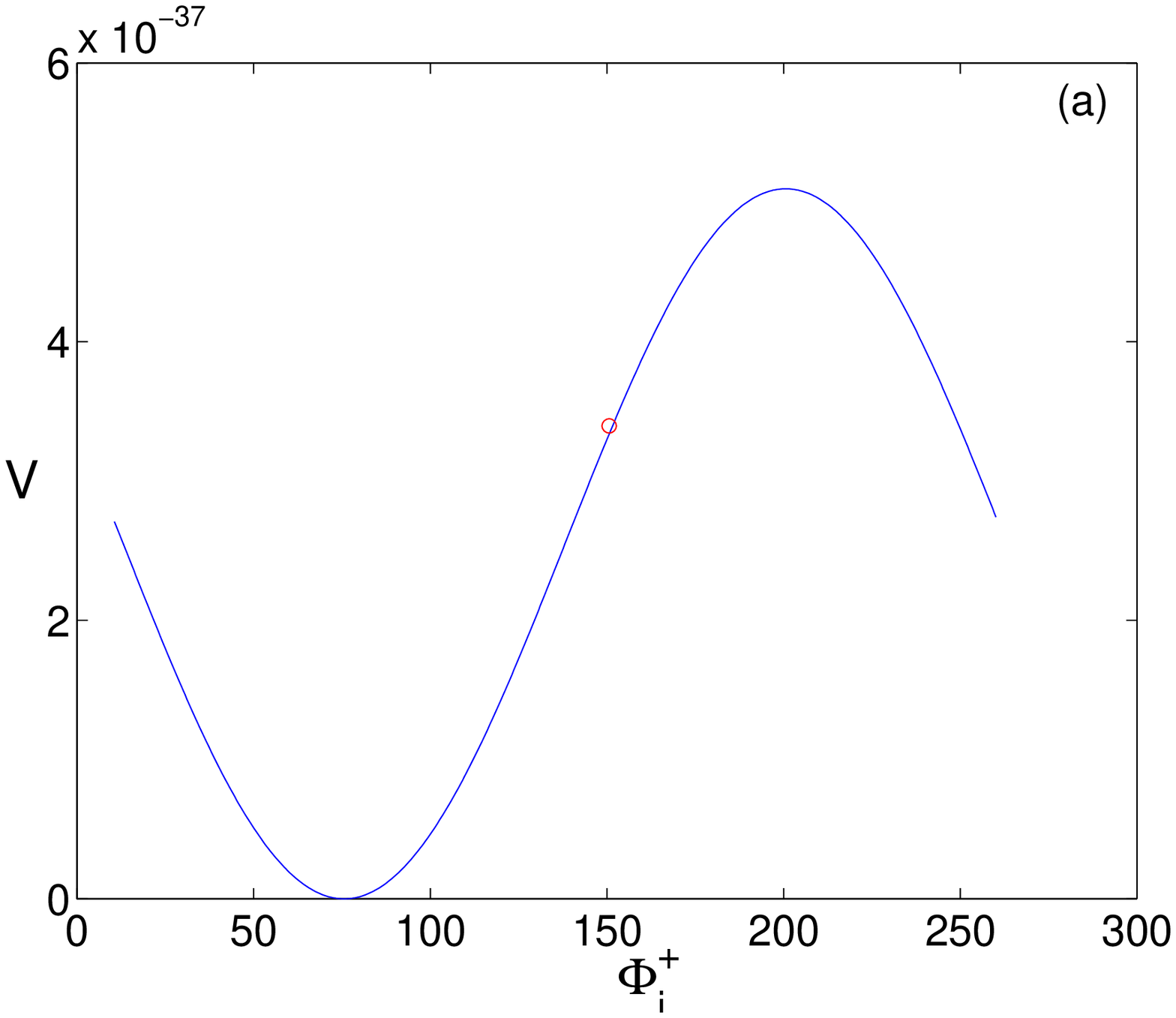}
\includegraphics[width=7.5cm]{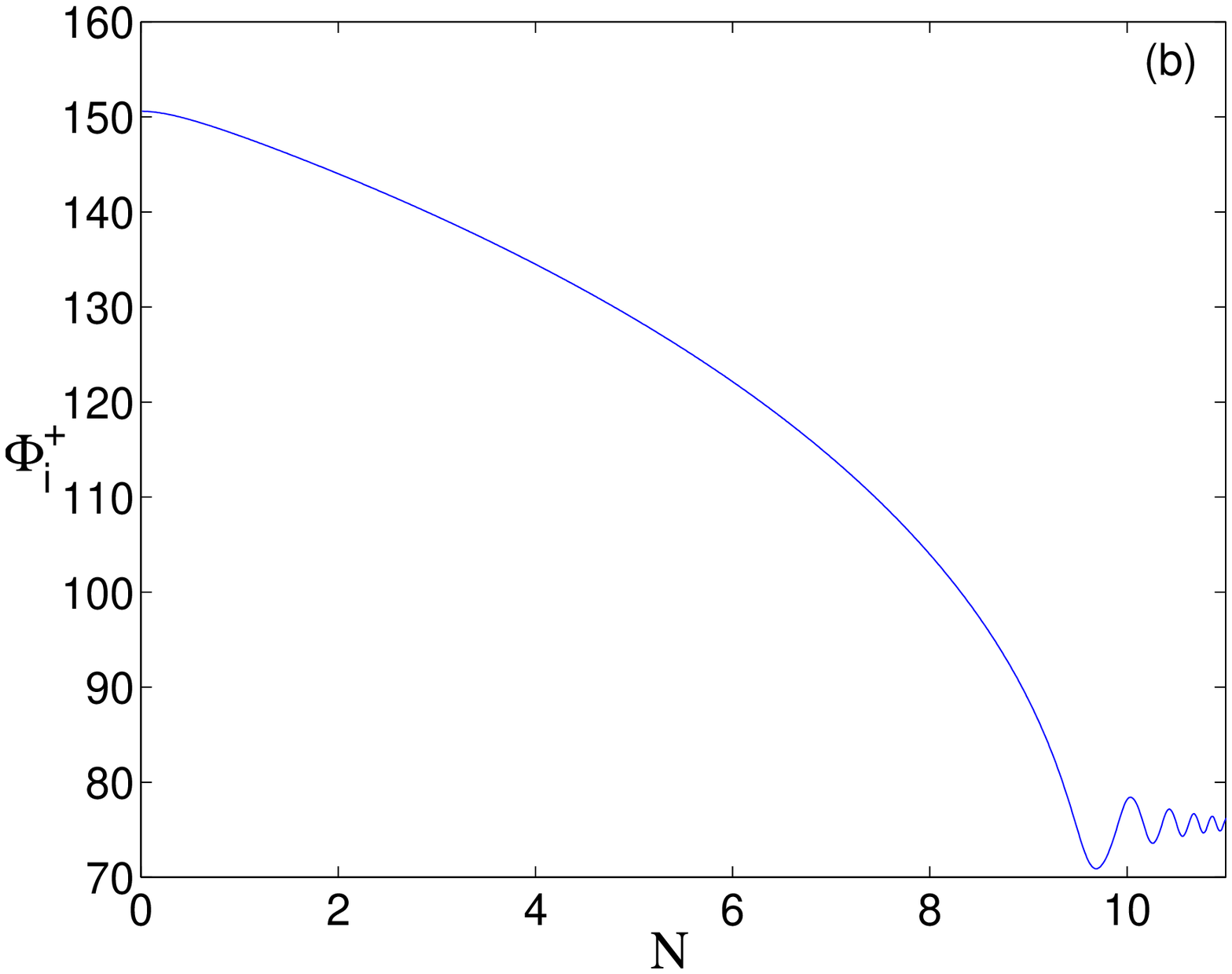}
\caption{ \textbf{\sffamily (a)} Plot of the potential $V$ as a function
of $\Phi^+_{\rm i}$ (in $M_{\rm P}$ units) along the bottom of the valley 
defined by $\Phi^-_{\rm i}={\rm constant}$, for the same example as in 
Fig.~1b, with $n_3=0.98$;
\textbf{\sffamily (b)} evolution of $\Phi^+_{\rm i}$ as a function of 
$N$ (the number of e-folds) along the same valley as before and initial 
conditions given by the circle shown in {\sffamily (a)}.}
\label{fig3}
}
}

As we saw in section~\ref{imdir}, with three condensates we now have 
the enticing possibility of lifting the flat direction in the imaginary 
fields $\Phi^+_{\rm i}$ with a sizeable effect. We will then be 
interested in the evolution of the field along this previously flat 
direction.

In order to illustrate the lifting of the flat direction, let us assume 
from now on that we have a perfectly well defined minimum with two 
condensates, i.e. we define $W_0$, given by Eq.~(\ref{sup2c}), as our 
starting superpotential. Either by K\"ahler stabilization (with $n_1=n_2$)  
or pure racetrack (in which $n_1 \neq n_2$ is required), the corresponding 
scalar potential provides us with minima for $\Phi_{\rm i}^-$ (given by 
Eq.~(\ref{mincond})) and $\Phi_{\rm r}^{\pm}$ (as explained in 
section~\ref{realdir}), whereas $\Phi_{\rm i}^+$, as defined in 
Eq.~(\ref{imflat}), is the flat direction.

Now let us add a third condensate to this system, in such a way that the
position of the already determined minimum is not substantially affected.
This will happen if the $\beta$ function coefficient of this third
condensate is considerably smaller than that of the other two, so that
$\alpha_3= 24 \pi^2/(3 N_3-M_3) \ll \alpha_1, \alpha_2$. This third
condensate, with gauge kinetic function given by $4 \pi f_3=S-n_3 T/2$, will
contribute to the scalar potential through a term in the superpotential given
by $C_3 e^{-\alpha_3 f_3}$. This will generate an extra piece in the scalar 
potential along the formerly flat direction given by 
\be
V_{\rm new} \approx V_{13} (1+\cos(\varphi_{13})) \;\;,
\label{Vcos}
\ee
where
\begin{eqnarray}
V_{13} & = & 2 e^K C_3 e^{-\alpha_3 (\sr-n_3 \tr/2)}\left[  
\frac{1}{K_{S \bar{S}}} (K_S W_0|_{\rm min}+ {W_{0S}|_{\rm min}}) 
\left( K_S-\frac{\alpha_3}{4 \pi} \right) \right. \nonumber \\
& + & \left. \frac{1}{K_{T \bar{T}}} (K_T W_0|_{\rm min}+{ W_{0T}|_{\rm min}}) 
\left( K_T+\frac{n_3 \alpha_3}{8 \pi} \right) - 3 W_0|_{\rm min} \right] \;\;,
\label{pref}
\end{eqnarray}
with $W_0|_{\rm min} = C_1 e^{-\alpha_1 (\sr-n_1 \tr/2)} - C_2 
e^{-\alpha_2 (\sr-n_2 \tr/2)}$, $\Phi^-_{\rm i}$ is fixed at the value for
which the two first condensates have opposite phases and
\be
\varphi_{13} = \frac{1}{4 \pi} \left[ (\alpha_3-\alpha_1) \si -
(n_3 \alpha_3 - n_1 \alpha_1) \frac{\ti}{2} \right] \;\;.
\label{angle}
\ee
In Fig.~3a we have plotted the potential $V_{\rm new}$ versus 
$\Phi_{\rm i}^+$ for the same example as in Fig.~1b, this time with 
$n_3=0.98$. The shape of this potential follows, as expected, 
Eq.~(\ref{Vcos}). Concerning the cosmological evolution of the now non-flat 
field $\Phi_{\rm i}^+$, this is shown in Fig.~3b for the same example as 
given in Fig.~3a with the initial condition there represented by a
circle. This corresponds to starting the evolution with
a potential energy equal to $60\%$ of the maximum of the cosine potential,
and it lasts for less than 10 e-folds.  
The model corresponds to the so-called natural inflation \cite{Freese90}.

Encouraged by these results, we will now try to find out under which conditions
we could achieve more e-folds of inflation. It turns out that the closer
$n_3$ is to $n_1$ the flatter the potential is along $\Phi_{\rm i}^+$. In 
fact for $n_3=n_1$ we should recover the flat direction that the two first
condensates had defined. Therefore imposing $n_3 \sim n_1$ should give
us a potential as flat as we wish. This is shown in Fig.~4a, where
we evolve the field $\Phi_{\rm i}^+$ for the same example as before, only
that this time we set $n_3=0.995$. As we can see, we can easily obtain 
more than 60 e-folds of inflation.

\FIGURE{
\parbox{15cm}{
\includegraphics[width=7.5cm]{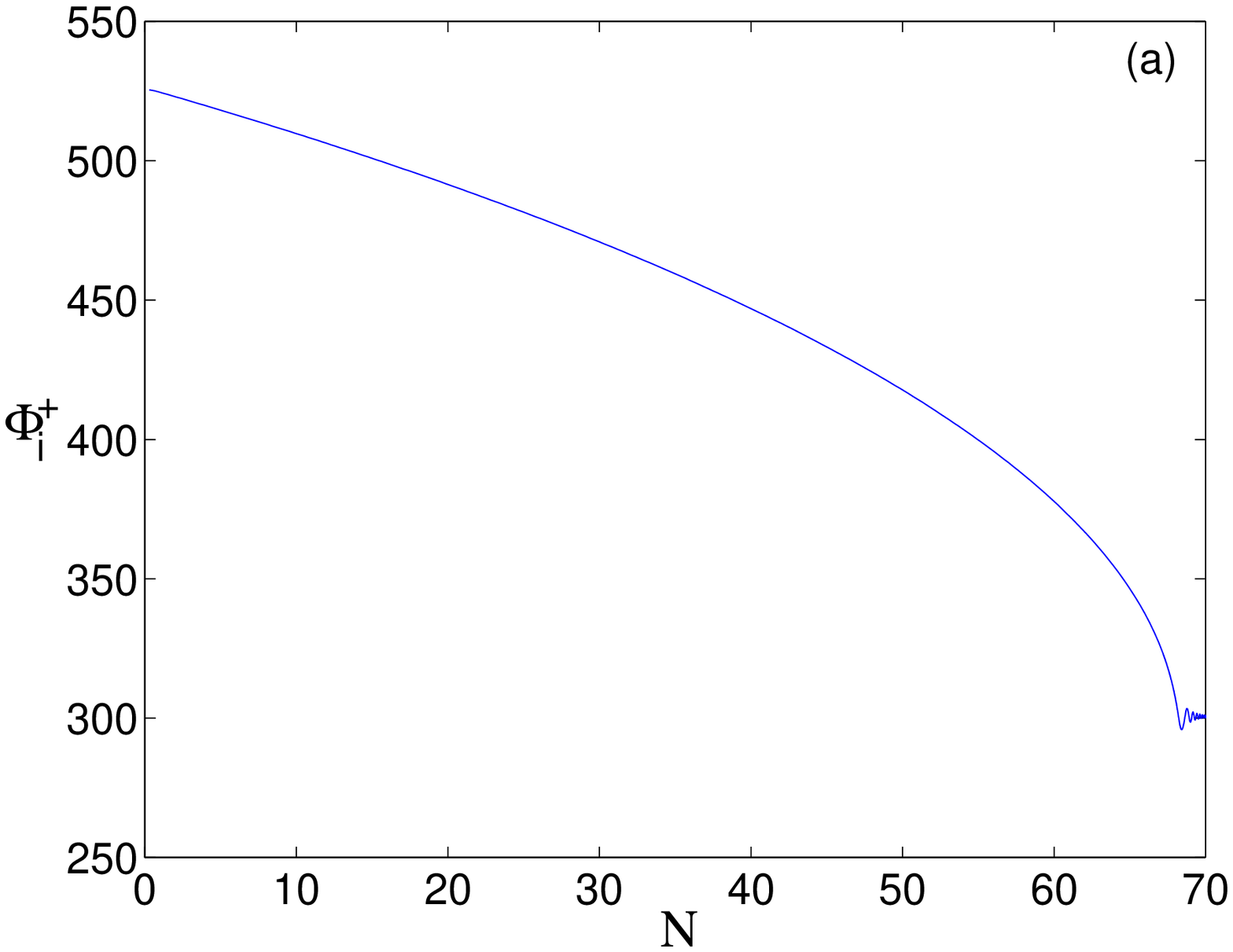}
\includegraphics[width=7.5cm]{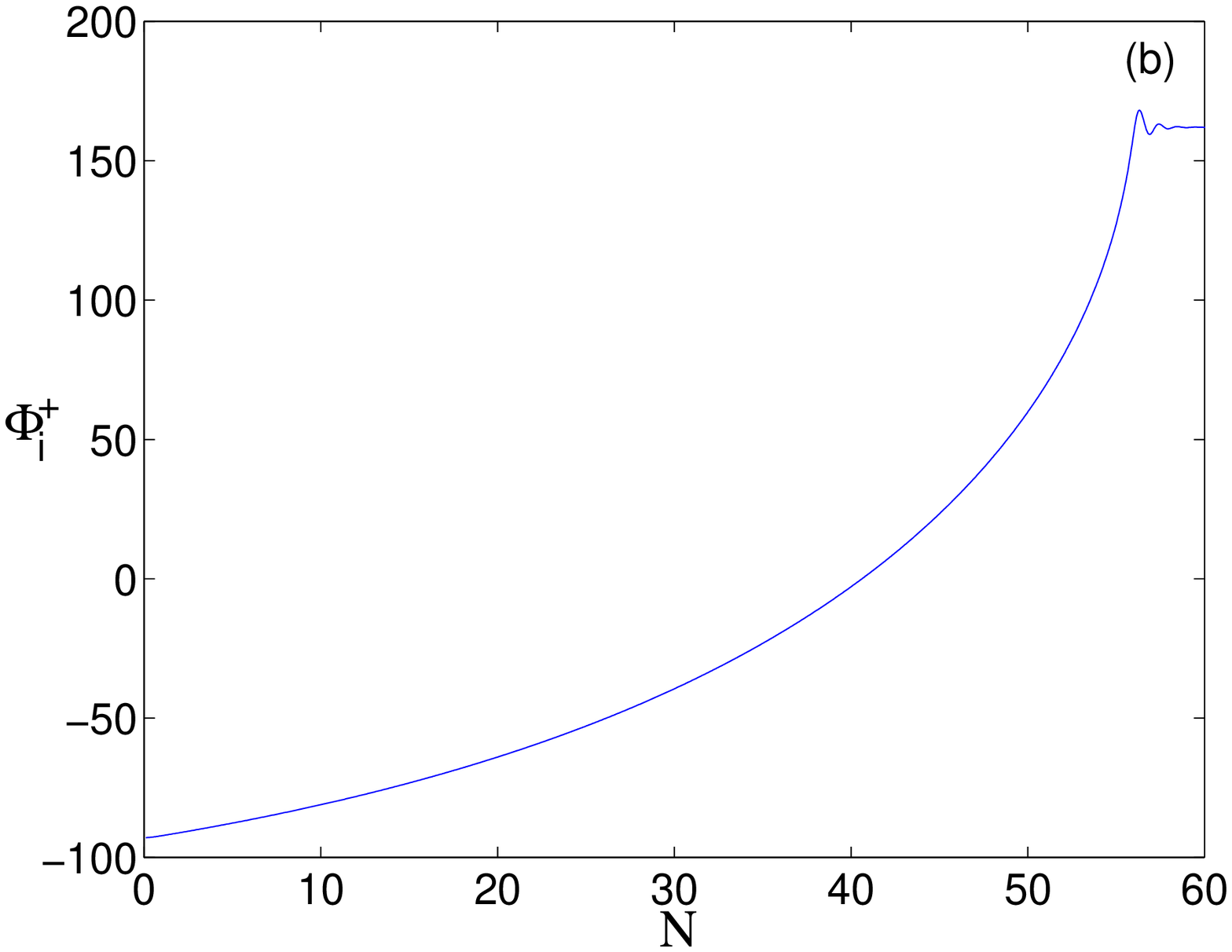}
\caption{Evolution of $\Phi^+_{\rm i}$ (in $M_{\rm P}$ 
units) as a function of $N$ (the number of e-folds) along the bottom of the 
valley defined by $\Phi^-_{\rm i}={\rm constant}$, for:
\textbf{\sffamily (a)} the same 
K\"ahler stabilization model as in Fig.~3, with $n_3=0.995$ and similar 
initial conditions;
\textbf{\sffamily (b)} a pure racetrack model given by $N_1=5$, 
$M_1=7$, $N_2=4$, $M_2=0$, $N_3=3$, $M_3=2$, $n_1=n_3=1$, $n_2=1.04$ and 
initial conditions near the maximum of its potential.}
\label{fig4}
}
}

The same is true for the pure racetrack case, which is shown in Fig.~4b.
The example we have chosen is the last one of Table~1, to which we
have added a third condensate given by $N_3=3$, $M_3=2$, $n_3=1$. In
general these models suffer from the problem that the phenomenologically
acceptable vacua correspond to potentials that are not flat enough. The 
closer we try to bring all three $n_i$ parameters to each other in order to 
create a smaller slope, the smaller the gravitino mass becomes. One of the 
best examples we have is the one shown here, where we obtain more than 
50 e-folds of inflation, but at the price of starting the evolution almost at 
the maximum of the potential. From this model-building point of view, the 
models with K\"ahler stabilization are somehow favoured.
Also note that, in order to perform the evolution of the fields in this
pure racetrack models, we have had to cancel the negative vacuum energy
by adding a constant to the scalar potential. This is equivalent to assuming
that some unknown mechanism is responsible for the cancellation of the
cosmological constant.

Finally let us add a few comments concerning the scale of these types of
inflation. The energy scale of the potential is similar to the scale of 
the gaugino condensates, that is around $10^{11} \; \gev$. As the model 
stands, a simple single-field slow-roll type of inflation, this would 
undoubtedly be too low  to explain the cosmic microwave background anisotropy 
and large scale structure. This flat direction could eventually be 
incorporated in a hybrid inflation type of scenario \cite{Copel94,linde91}, 
where the density fluctuations can be generated with this sort of energy 
scale. This, however, would go far beyond the scope of this paper.
Alternatively, this flat direction could be used to implement a scenario of 
weak inflation \cite{Randa95}, of the type used to solve the moduli
problem.

\section{Moduli problem}

We are going to conclude this analysis of the cosmological behaviour of 
the moduli in heterotic M-theory by revisiting the moduli problem from 
that point of view. Moduli are, in the weakly coupled heterotic string, 
very dangerous relics from the cosmological point of view, given that they
have large vacuum expectation values (of the order of $M_{\rm P}$) and 
light masses (of the order of the gravitino mass, $m_{3/2}$). The potential
cosmological problems of light, weakly interacting particles had
already been pointed out in the context of supergravity \cite{Cough83}
and superstrings \cite{Ellis86}, but were explicitly addressed for moduli
in Refs.~\cite{Decar93p}. Altogether, if moduli decay they should do so 
before nucleosynthesis, in order not to spoil its predictions for the 
abundance of light elements. This imposes a lower bound of their masses
given by
\be
m_{S, T} \leq O (10 \; {\rm TeV}) \;.
\label{lowerb}
\ee
On the other hand, if these particles do not decay they should not
overclose the Universe today. This imposes an upper bound on their
masses, in the absence of inflation, given by $m_{S, T} \geq  1 \; 
{\rm keV}$. This bound applies both to the moduli and their fermionic 
partners, the modulini; in principle, it could be avoided with a 
period of inflation at a scale low enough to dilute them and avoid their
regeneration in large amounts. 

However, for the scalars there is an additional problem, which is
that, at the end of inflation, these particles will probably be shifted 
from their zero temperature minima. It has been estimated that during
their evolution towards their true minima, these fields will oscillate
and their energy density will increase substantially. For stable moduli
this imposes an upper bound on their masses given by
\be
\frac{m_{\phi}}{100 \; {\rm GeV}} \leq \left( 10^{10} 
\frac{\phi_{\rm in}}{M_{\rm P}} \right)^{-4} \;\;,
\label{osc}
\ee
where $\phi=S,T$ and $\phi_{\rm in}$ is the value of the shift at the end of 
inflation, which is typically of order $M_{\rm P}$ (actually, in the examples
presented above it is of order $100 \;M_{\rm P}$, see Figs.~3 and 4). From 
here we get extremely tight upper bounds for the masses of the stable moduli, 
of the order of $10^{-24} \; {\rm eV}$.

Several mechanisms have been proposed to cure this problem, the most
promising of which is to invoke a period of thermal inflation 
\cite{Lyth95} just before nucleosynthesis in order to dilute these 
unwanted relics. In particular, it has been shown in Ref.~\cite{Asaka99} 
that it can solve the overclosure problem of the stable moduli if their 
masses are in the range of $10 \; {\rm eV}$ to $10^{4} \; {\rm GeV}$. 
This is particularly important for the forthcoming discussion.

Let us then turn to the issue of moduli masses in these heterotic
M-theory models we have just analysed. We have calculated and
diagonalized the mass matrix for the four real components, namely
$\sr$, $\tr$, $\si$ and $\ti$, both for the cases of K\"ahler stabilization
and pure racetrack models. In both scenarios the results are rather similar
and the eigenvectors do correspond to the directions $\Phi^+_{\rm r,i}$,
$\Phi^-_{\rm r,i}$ defined in previous sections, which shows that our 
choice corresponds to the physical directions. To illustrate this, we
present two examples where we have calculated the moduli masses.
\begin{itemize}
\item $K_{\rm np} \neq 0$. The example chosen is 
${\rm SU}(3)_{M=0} \times {\rm SU}(4)_{M=8} \times {\rm SU}(2)_{M=0}$, with 
$n_1=n_2=1$
\begin{eqnarray}
m_{\Phi_{\rm r}^-}, m_{\Phi_{\rm i}^-}  & \sim & 3 \times 10^3  \; m_{3/2} 
\nonumber  \\
m_{\Phi_{\rm r}^+} & \sim & 40 \;  m_{3/2}  
\label{massKnp}
\end{eqnarray}
\begin{center}
\begin{tabular}{lcccc}
$n_3=0.98$ & $\rightarrow$ & $m_{\Phi_{\rm i}^+}$  & $\sim$ & $10^{-3} \; 
m_{3/2}$  \\
$n_3=0.995$ & $\rightarrow$ & $m_{\Phi_{\rm i}^+}$  & $\sim$ & $10^{-4} \; 
m_{3/2}$  
\end{tabular}
\end{center}
\item $K_{\rm np} = 0$. The example chosen is 
${\rm SU}(3)_{M=2}\times {\rm SU}(5)_{M=7}\times {\rm SU}(4)_{M=0}$, with 
$n_1=n_2=1$
\begin{eqnarray}
m_{\Phi_{\rm r}^-}, m_{\Phi_{\rm i}^-}  & \sim & 10^4 \; m_{3/2} \nonumber \\
m_{\Phi_{\rm r}^+} & \sim & 2 \; m_{3/2}  
\label{massK0}
\end{eqnarray}
\begin{center}
\begin{tabular}{lcccc}
$n_3=1.04$ & $\rightarrow$ & $m_{\Phi_{\rm i}^+}$  & $\sim$ & $7 \times 
10^{-4} \; m_{3/2}$ \\
$n_3=1.05$ & $\rightarrow$ & $m_{\Phi_{\rm i}^+}$  & $\sim$ & $6 \times 
10^{-4} \; m_{3/2} \;\;.$
\end{tabular}
\end{center}
\end{itemize}
As we can see, the combination of moduli associated with the steepest
directions, $\Phi^-_{\rm r,i}$, are heavy enough to fulfil the lower bound
given by Eq.~(\ref{lowerb}), and it is the other two combinations of 
moduli, $\Phi^+_{\rm r,i}$, which are in trouble with the existing bounds. 
In particular, the mass associated with the almost flat direction 
$\Phi^+_{\rm i}$ is, as expected, smaller as $n_3$ is closer to either 
$n_1$ or $n_2$; it should be exactly zero in the limit in which it becomes 
equal to one of them, i.e.\ when we recover the flat direction typical of the 
two-condensate case. Such a light modulus is not likely to decay at all,
and therefore the bound given by Eq.~(\ref{osc}) should be applicable to 
it. A period of thermal inflation would then be desirable for these two 
lighter moduli not to become unwanted relics.

We have also analysed the possible reasons for this pattern of masses 
to be so different from that of the weakly coupled heterotic models, in
which all moduli masses were of the order of $m_{3/2}$. Apart from the
different curvatures of the potential along the different directions,
which are now different and explain the existing hierarchy between the 
different combinations of moduli, the difference in magnitude can be
attributed to the difference of the magnitude of the K\"ahler metric at the
minimum of the potential. As we know, and as was pointed out in the 
previous section, the moduli $S$ and $T$ are not canonically normalized 
particles as the kinetic terms of their Lagrangian are given by
\be
{\cal L}_{\rm kin} = K_{S \bar{S}} D_{\mu} S D^{\mu} \bar{S} +  
 K_{T \bar{T}} D_{\mu} T D^{\mu} \bar{T} \;\;.
\label{kin}
\ee
In order to obtain the physical masses, we must transform the fields
into canonically normalized ones, and the transformation induces
inverse powers of $K_{S \bar{S}}$, $K_{T \bar{T}}$ in the mass matrix.
So far, this was also the case in the weakly coupled heterotic string,
the difference being the different magnitudes of these derivatives
in the two theories. Before, the expected values for the moduli at the
minimum were or order 1, and therefore these normalization factors 
would also have the same size (note that, for example, for $K_0$ we have
$K_{S \bar{S}} \sim 1/(2 \sr)^2$, $K_{T \bar{T}} \sim 3/(2 \tr)^2$).
However, in heterotic M-theory we look for minima of the potential
corresponding to $S$, $T$ of order 20 (always in $M_{\rm P}$ units), which
implies that the normalized masses are suppressed by factors of
$10^{-4}$-$10^{-3}$, and therefore enhanced in absolute value.

\section{Conclusions}

In this paper we have studied the dynamics of the two typical M-theory
moduli, namely the volume of the 6-dimensional manifold $V$ and
the orbifold that parametrizes the $11^{\rm th}$ segment $\pi \rho$ in
terms of the dilaton $S$ and an overall modulus $T$.
Assuming gaugino condensation in the hidden wall as the source of SUSY 
breaking, we studied the resulting scalar potential for different numbers 
of condensates in the case of the so-called K\"ahler stabilization (where 
$K=K_0+K_{\rm np}$) and pure racetrack models ($K=K_0$). In both cases a 
determinant feature of the vacuum structure was the field dependence of
the superpotential and scalar potential, essentially given in terms of 
the gauge kinetic functions for each condensing group, $4 \pi f_a = S-n_a
T/2$.  The interplay between condensates therefore fixes a particular 
combination of $S$ and $T$ (which we denoted by $\Phi^-_{\rm r,i}$), the 
orthogonal ones ($\Phi^+_{\rm r,i}$) being potentially flat. While the 
potential flatness of $\Phi^+_{\rm r}$ is lifted by the presence of the 
K\"ahler potential, we have shown that indeed with one and two condensates 
$\Phi^+_{\rm i}$ is always flat, whereas for three or more condensates the 
flatness is preserved if all the gauge kinetic function parameters $n_i$ 
are the same.

After finding different examples of condensing groups leading to 
phenomenologically acceptable vacua, we proceeded to study the dynamical
behaviour of these four combinations of moduli. While $\Phi^-_{\rm r,i}$
behave very much like the dilaton $S$ of the weakly coupled heterotic 
string, therefore are not suitable to be inflatons, we solved the evolution
equations in an expanding Universe for the flatter directions given by
$\Phi^+_{\rm r,i}$. With two condensates the potential along $\Phi^+_{\rm r}$
behaves as an inverse power-law, which would be inflating for long 
enough time if it weren't for the fact that the field is not canonically 
normalized. As it stands, we only get a few e-folds of intermediate 
inflation. However, having lifted the flat direction $\Phi^+_{\rm i}$ with
a third condensate, we have shown how to achieve a flat enough cosine
potential for it, which leads to more than 60 e-folds of natural inflation.
Therefore the axion $\Phi^+_{\rm i}$ becomes a suitable inflaton in heterotic
M-theory.

Finally we have considered the status of the moduli problem in these models.
While the steeper directions $\Phi^-_{\rm r,i}$ have associated masses well
above the bound of $10 \; {\rm TeV}$ required to solve the moduli problem
for decaying particles, the flatter directions are too light to save it,  
and also above the existing upper bound for stable particles. Therefore
a period of low-temperature thermal inflation would be required to dilute 
them. Concerning their magnitude, the size of the K\"ahler metric at the 
minimum of the potential is responsible for the enhancement in magnitude
of these masses with respect to those characteristic of the weakly coupled 
heterotic case.

\vspace{1cm}

\acknowledgments

It is always a pleasure to thank Alberto Casas for his invaluable
advice and continuous encouragement. We also thank K.~Choi, H.B.~Kim, 
G.~Lopes~Cardoso, A.~Lukas and S.~Stieberger for discussions, and 
J.~Generowicz for his help with computer related questions.
TB thanks the CERN Theory Division for hospitality during the initial 
stages of this project.


\end{document}